\documentclass[12pt]{article}
\usepackage{graphicx}
\usepackage{amssymb,amsmath}
\usepackage{url}

\begin{document}

\date{}

%\title{- Letter of Intent - \\[3ex]
%\bf Planck-scale gravity test at PETRA}

\title{ {\bf Planck-scale gravity test at PETRA} \\[3ex]
Letter Of Intent}

\author{V.Gharibyan\thanks{Electronic address: vahagn.gharibyan@desy.de}, K.Balewski 
 \\ [1ex]
\small\it{Deutsches Elektronen-Synchrotron DESY,}  \\
\small\it{Notkestrasse 85, D-22603 Hamburg, Germany }
}

\maketitle

\begin{abstract}
 Quantum or torsion gravity models predict unusual properties of space-time at very 
short distances. 
In particular, near the Planck length, around $10^{-35}m$, empty space 
may behave as a crystal, singly or doubly refractive. This hypothesis, however, 
remains uncheckable for any direct measurement since the smallest distance 
accessible in experiment is about $10^{-19}m$ at the LHC. Here we propose a 
laboratory test to measure space birefringence or refractivity induced
by  gravity. A sensitivity $10^{-31}m$ for doubly and $10^{-28}m$ for singly 
refractive vacuum could be reached with PETRA 6~GeV beam exploring UV laser Compton 
scattering. 
\end{abstract}

\newpage

\tableofcontents
\newpage

\section {Introduction}

The quantum formalism can not be directly applied to gravitation
and that is one of the major problems on a way of understanding and describing 
the physical reality.
An important reason for this is the dynamical space concept adopted in
general relativity, the currently accepted theory of gravity which states that any 
mass or particle modifies the space geometry (or metrics). On the other hand, the successful 
quantum theories within the Standard Model operate only in a fixed geometry space. 
For instance, observed violations of the discrete symmetries such as space, charge and
time parities are attributed to the particles and their interactions while the scene 
of the interactions, the space-time, is considered to remain perfectly 
symmetric~\cite{Amsler:2008zzb}. These two faces of space 
are believed to unify at distances near the Planck length
$l_P =1.6\cdot 10^{-35}m$  (or mass $M_P = 1.2\cdot 10^{19}GeV$, natural units are 
assumed throughout the letter). 
At this scale gravity is expected to be similar in strength to the
electroweak and strong forces and quantum effects become important for
the gravitational field. String theory and loop quantum gravity
theory are prominent candidates which set a framework to make
predictions in that energy domain. In many cases, unconventional
space-time properties are suggested, such as vacuum 
refractivity~\cite{AmelinoCamelia:1997gz}  and/or 
birefringence~\cite{Gambini:1998it}. 

Such effects may be studied by using lasers and high energy accelerator beams as recommended in 
ref.~\cite{Gharibyan:2012gp}. 
The proposed experiment at PETRA will probe the vacuum symmetry in a search for a handedness or 
chirality of the empty space presumed by quantum gravity.
A figure of merit is circular birefringence \hbox{$\Delta n = n_L - n_R$} of 
space, with $n_{L(R)}$ being the refraction index of left(right) helicity photons 
traversing the space. Average refraction \hbox{$n = (n_L + n_R)/2$} will be tested with additional instrumentation. 

In the following, quantitative theoretical estimates and existing experimental limits are 
quoted, the formalism of the suggested method is presented, and the
proposed experimental setup is described. Expected performance, experimental reach
with statistical and systematic accuracy estimates are discussed as well. 

\section {Photon dispersion at Planck-scale}

Since Planck mass $M_P = \sqrt{c \hbar /G}$ is build from the speed of light and fundamental 
Planck and gravitational constants, this mass scale is considered to be relativistic 
and quantum gravitational. 
Most general modification of photon dispersion relation at lowest order of 
Planck mass could be expressed as
\begin{equation}
\omega^2 = k^2 \pm \xi \frac{k^3}{M_P}
\label{eq3}
\end{equation}
where $\omega$ and $k$ are photon's energy, momentum, respectively,  
while the $\xi$ is a dimensionless parameter and the $\pm$ signs stand for 
opposite helicity photons. This is main relation we are going to test at PETRA.

Several theories are predicting or supporting the relation (\ref{eq3}). 
The Planck scale quantum gravity modifies the Maxwell equations by 
adding extra terms proportional to the Planck length~\cite{Gleiser:2003fa}:
\begin{equation}
\frac{\partial{\vec{E}}}{\partial{t}} = \vec{\nabla}\times\vec{B}-2\xi l_{P} 
\vec{\nabla}^{2} \vec{B}
\label{eq1}
\end{equation}
\begin{equation}
\frac{\partial{\vec{B}}}{\partial{t}} = -\vec{\nabla}\times\vec{E}-2\xi l_{P} 
\vec{\nabla}^{2} \vec{E}
\label{eq2}
\end{equation}
which leads to a deformed energy-momentum or dispersion relation (\ref{eq3}).
In the above equations,  $\vec{E}$ and $\vec{B}$  describe the electromagnetic field.
More general expressions accounting for space anisotropy are derived 
in Ref.~\cite{Gubitosi:2010dj}. 
Using conventional definition $n=d\omega /dk$,
it is easy to verify that Eqs. (\ref{eq3})-(\ref{eq2})  
introduce a chiral vacuum with an energy dependent  birefringence
\begin{equation}
\Delta n = 3\cdot 10^{-19} \cdot \xi \cdot \omega [GeV]
\label{eq4}
\end{equation}
where the magnitude of $\xi$ defines the characteristic energies or distances where
quantum-gravity effects become sizeable. 
In the simplest possible picture, this only happens at the Planck
scale, and hence $\xi=1$. However, the running of fundamental
constants with energy may require quantum gravity to become active a
few orders of magnitude below the Planck scale. The parameter $\xi$ is
there to account for such effects.

Another possible source of vacuum chirality is described by 
torsion gravity, an extension of the general relativity into the microscopic 
world to include particles' spins - for a review see~\cite{Hehl:1976kj}. 
In general, the spin gravity (space torsion) is considered to be weaker than the 
mass gravity (space curvature). However, near the Planck scale it may become detectable. 
Following Ref.~\cite{Prasanna:2009zz}, from the electromagnetic field Lagrangian 
\begin{equation}
{\mathcal L} =-\frac{1}{4}F_{\mu\nu}F^{\mu\nu}+qT^{\mu\nu\rho}
(\partial^\sigma F_{\mu\nu})F_{\rho\sigma}
\label{eq5}
\end{equation}
with a torsion tensor $T^{\mu\nu\rho}$ and free parameter $q$ one derives a dispersion 
relation quite similar to Eq.(\ref{eq3}) 
\begin{equation}
\omega^2 = k^2 \pm q S_0 k^3 
\label{eq6}
\end{equation}
where $S_0$ stands for a time component of the contorsion vector.

Myers and Pospelov~\cite{Myers:2003fd} derived the expression~(\ref{eq3}) 
within effective field theory with dimension 5 operators. 
A similar effect is calculated in ref.~\cite{Dalvit:2000ay} exploring graviton 
interaction with electromagnetic field in one-loop approximation. In our setup
the gravitons emerge from the gravitational field of Earth.
In summary, chiral space is a universal feature of Planck-scale gravity, in
the sense that it is predicted by a large diversity of theories.

A nonbirefringent gravitational space is also possible
and has been  predicted within String theories using D-brane formalism. 
In Ref.~\cite{Ellis:2008gg} a polarization independent 
refractivity 
\begin{equation}
 n-1=\zeta \frac{k}{M_P}
\label{eq7}
\end{equation}
is obtained for the space-time foam near the Planck length.
Here we use $\zeta$  instead of the $\xi$ to distinguish between the nonchiral and 
chiral space. In principle, both types may occur in the same vacuum at different scales  
$\zeta$ and $\xi$.
Both gravity induced effects, namely birefringence and
refractivity, share the common feature that their strength is growing
with the photon energy. This is in contrast to
the usual condensed matter or electromagnetic,
nontrivial vacua where the refraction effects are suppressed by powers of 
the energy~\cite{Dittrich:1998fy,Bombelli:2004tq}. 
Such growth should allow to approach Planck scale at the PETRA as will be shown below..  

\section{Current limits}

Experimental limits on space chirality are set by astrophysical observations  
exploring  birefringence induced depolarization of the linear light which comes from 
distant cosmological sources~\cite{Gleiser:2001rm}. The limits, however, are based on 
assumptions about the origin, spatial or temporal distribution
of the initial photons, and their possible interactions during the travel.  
Another critical assumption is a uniformly distributed birefringence over cosmological distances.
The most stringent limit $\xi < 2.4\cdot 10^{-15}$ is set 
by Ref.~\cite{Stecker:2011ps} based on photons with polarization $0.63\pm0.30$ in an energy 
range from 100 to 350~keV from GRB041219a~\cite{Glynn:2007}.
Sensitive particle-physics effects have been suggested to test
quantum gravity, mainly using threshold energies~\cite{Heyman:2003hs}.
Applying cosmic ray constraints on photon decay and vacuum Cherenkov 
radiation~\cite{Gharibyan:2003fe}, one arrives to $\zeta < 30$ and 
$\zeta < 300$ limits, respectively.

For the space refractivity, there are astrophysical observations interpreted~\cite{Ellis:2009yx} 
as $\zeta \sim 10$. This
is derived from energy dependent time delay measurements of  photons 
from  distant sources. 
Similar to the results derived from polarized photons of
cosmological origin, strong assumptions have to be made on the source
of these photons.
In addition,  quoted astrophysical constraints are 
valid only for photon-virtual graviton loop interactions, since the photon path
is essentially free from gravitational fields.   

PETRA measurements could shed light on the quantum-gravity
promoted space chirality and refractivity including effects introduced by 
Earth gravitons.
 In the laboratory the Planck scale
can be accessed by exploring the extreme sensitivity of the high energy Compton 
scattering to the vacuum refraction as discussed in the following.

\section{Compton scattering affected by gravity}

\begin{figure}
\centering
\includegraphics[scale=0.50]{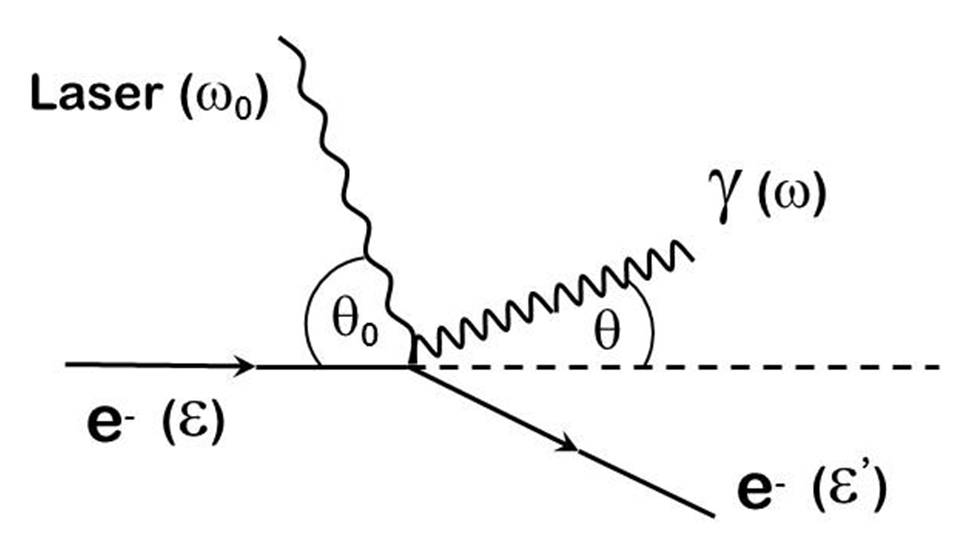}
\caption{\label{comp-diag}
Compton scattering schematics.}
\end{figure}

Let us denote by $\omega_0$, $\omega$, $\theta_0$, $\theta$ the energies and angles of the
incident and scattered photons relative to the initial electron direction as illustrated in 
Figure~\ref{comp-diag}.
Then, according to Ref.~\cite{Gharibyan:2003fe}, for the high energy Compton 
scattering in a vacuum with $n\approx 1$ (up to $\mathcal{O}((n-1)^2)$ terms),
the energy-momentum conservation yields
\begin{equation}
   n-1=\frac{\mathcal E}{2\gamma^2 ({\mathcal E}-\omega)}\Biggl(
   1+x+\theta^2\gamma^2-x\frac{\mathcal E}{\omega}\Biggr)
\label{eq8}
\end{equation}
where $\gamma$,$\mathcal{E}$ are the Lorentz factor and energy of the initial
electron, \hbox{$x\equiv 4\gamma \omega_0\sin^2{(\theta_0/2)}/m$},
and $n$ is the refraction index for the direction $\theta$ and energy $\omega$.
This formula is more general than  Eq.(3) of Ref.~\cite{Gharibyan:2003fe}.
The difference is in a factor ${\cal E}/({\cal E}-\omega)$, because
in contrast to \cite{Gharibyan:2003fe} the 
final photon mass squared \hbox{$k_\mu^2 = \omega^2 (1-n^2)$} is not neglected
for this Letter.

Substituting $n-1$ in the Eq.(\ref{eq8}) by the gravitational refractivity from the Eq.(\ref{eq7}) 
we can estimate how the quantum gravity would change the scattered photons' maximal energy 
$\omega_{m}$ (Compton edge, at $\theta=0$).
The expected shift of the Compton edge is
\begin{equation}
\Delta \omega_m \equiv
\omega_m (n)-\omega_m (1)=\frac{32\gamma^6 \omega_0^2\sin^4{(\theta_0/2)}}{(1+x)^4} 
\frac{\zeta}{M_P}
\label{eq9}
\end{equation}
relative to the vacuum (n=1) kinematics.
For optical lasers and head-on collision the kinematic factor $x\approx 2\cdot 10^{-5}\gamma$
and the right-hand side of Eq.(\ref{eq9}) grows as $\gamma^6$ at $GeV$ energies 
slowing down to $\gamma^2$ growth above $TeV$ energies.  
At sufficiently high $\gamma$, the huge value of  $M_P$ is compensated and the energy 
shift becomes detectable. 
Hence, this effect allows quantum-gravity 
induced space refractivity to be measured at PETRA by laser Compton scattering off 
electrons with $\gamma=11742$. 

\begin{figure}[htb]
\centering
\includegraphics[scale=0.50]{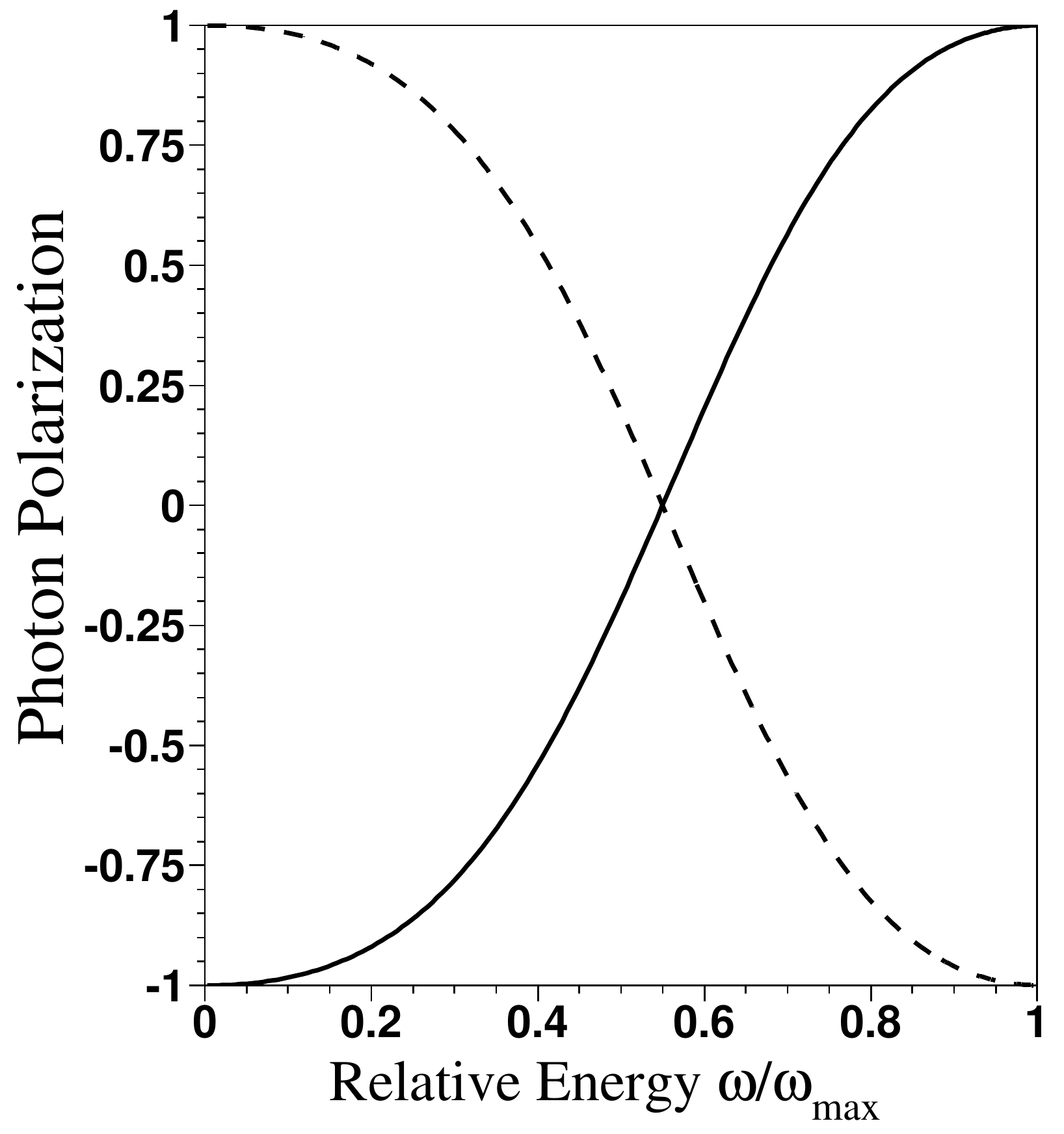}
\caption{\label{fig1}
Polarization of the Compton scattered photon on a 6~GeV electron as a function of 
the photon energy.
The solid and dotted lines correspond to the initial laser light helicity: +1 solid, -1 dotted.}
\end{figure}

In order to probe space birefringence, one needs to measure the
refractivity in Eq.(\ref{eq8}) for scattered photons of opposite helicity. 
This may be achieved by exploring circularly polarized initial laser beams and helicity 
conservation.  
The polarization of the secondary photons in the case of  scattering on
 unpolarized electrons is shown in Fig.~\ref{fig1},  using formulas from
Ref.~\cite{Lipps,mcmaster}.
At $\omega=\omega_{m}$ the polarization transfer is complete, such  
that the helicity of the Compton edge photons is fully defined by the laser light helicity. 
Consequently, in a birefringent vacuum the Compton edge energy is laser helicity dependent.
Evaluating Eq.(\ref{eq8}) for left and right helicity photons at $\theta=0$ yields
\begin{equation}
\Delta n = n_L (\omega_m^L) -  n_R(\omega_m^R) = \frac{(1+x)^2}{\gamma^2}A
\label{eq10}
\end{equation}
where $\omega_m^L$ and $\omega_m^R$ are the highest energies for the Compton  
opposite helicity photons and  \\
\hbox{$A=( \omega_m^L - \omega_m^R ) / ( \omega_m^L + \omega_m^R )$}
is an energy asymmetry.
 
Combining Eq.(\ref{eq10}) with the gravitational birefringence from the Eq.(\ref{eq4}) 
we arrive to
\begin{equation}
A = \frac{8\gamma^4 \omega_0 \sin^2{(\theta_0/2)}}{(1+x)^3} 
\frac{\xi}{M_P}
\label{eq11}
\end{equation}
which proves that for PETRA values of $\gamma$ the Planck scale space  
birefringence generates a measurable asymmetry.

In contrast to the astrophysical methods,  an accelerator Compton 
experiment is sensitive to the local properties of space at the laser-electron 
interaction point and along the scattered photon direction.
Hence, space isotropy tests are also possible as the accelerator rotates together 
with Earth. For any preferred direction the measured  birefringence
is expected to change as the scattered photon beam sweeps a circle over the celestial sphere.
For a given direction ($\delta$,$\alpha$) of the photon beam and a possible anisotropy 
axis ($\delta_0$,$\alpha_0$) one expects  
\begin{equation}
\Delta n = \Delta n_0 ( \cos{\delta} \cos{\delta_0}\cos{(\alpha-\alpha_0)}+\sin{\delta} \sin{\delta_0})
\label{eq12}
\end{equation}
where $\Delta n_0$ is the maximal birefringence, along the 
declination $\delta_0$ and right ascension $\alpha_0$. 
Despite of  tight limits set by low energy high precision experiments on 
space anisotropy~\cite{Sudarsky:2002ue}  the accelerator isotropy test 
is a valuable and complementary test at high energies.

\section{Proposed experiment}

\begin{figure}[h]
\centering
\includegraphics[scale=0.50]{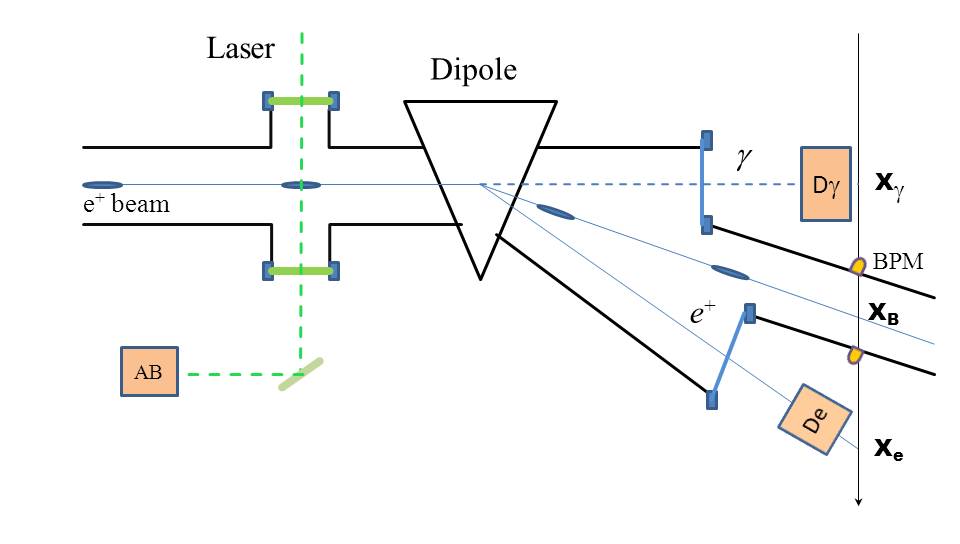}
\caption{\label{gscheme}
Schematic overview of the experiment.}
\end{figure}

In order to measure space birefringence and refractivity we propose a laser Compton 
experiment to be performed at the PETRA SW section.
The experiment will bring into collision PETRA bunches and light pulses from 
a mode locked laser to produce Compton photons. Scattered photons and positrons,
separated by a beam dipole magnet, are registered by downstream detectors.  
Positions and energies of the scattered secondaries will mainly be measured
in single particle detection mode. Positron beam position is measured using 
PETRA high precision BPM (Beam Position Monitor) system~\cite{Marx:2011zz}.
Laser beam intensity and polarisation will be monitored in a light Analyzer Box.
A schematic arrangement of the proposed experiment is presented in Figure~\ref{gscheme}.
Measured positions $X_e$ at laser light opposite helicities will allow to derive
space birefringence while the refractivity could be accessed using in addition 
horizontal positions of the beam $X_B$ and the Compton photon $X_\gamma$. 

We plan to run the experiment in 'parasitic' mode without disturbing 
user operations or affecting machine beam quality.

\subsection {Accelerator}

PETRA III~\cite{Balewski:2004iz}  is a third generation light source with 
6 GeV high quality positron beam. Main operational parameters of the 
machine~\cite{p3param} are collected in the table~\ref{tab1}. 
 
\begin{table}[!htb]
\caption{PETRA III parameters.}
\label{tab1}
\begin{center}
\begin{tabular}{|l|l|}
\hline\hline
\multicolumn{2}{|c||}{ \includegraphics[scale=0.8]{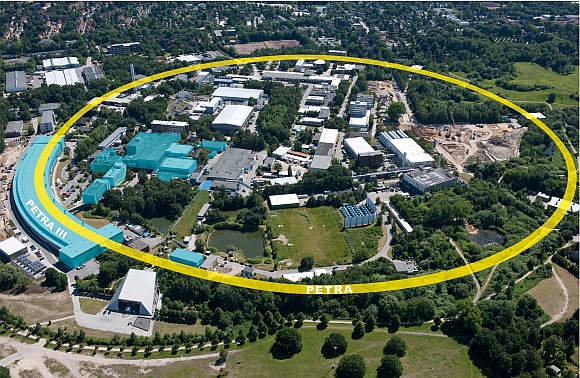} } \\
\hline
Positron energy & 6.0 GeV \\\hline
Circumference of the storage ring & 2304 m \\\hline
Harmonic number (buckets) & 3840\\\hline
Number of bunches &  40 -- 960 \\\hline
Bunch separation &  192 ns -- 8 ns\\\hline
Positron beam current & 100 mA (top-up mode)\\\hline
Horizontal  emittance & 1 $nm\cdot rad$ (rms)\\\hline
Vertical  emittance & 0.01 $nm\cdot rad$ (rms)\\\hline
Positron beam energy spread & 0.1\% (rms)\\\hline
RF & 499.564 MHz\\\hline
Revolution Time & 7.685 $\mu$sec \\\hline
Revolution Frequency & 130.121 kHz \\\hline
Bunch Length (rms) & 44 psec \\\hline
Positron Energy Loss per Turn from Dipoles & 1 MeV \\\hline
Overall Positron Energy Loss per Turn & 6 MeV \\\hline
%Positron Beam Lifetime (Multi Bunch Mode) & 10 h \\\hline
Positron Beam Lifetime (Time resolved Mode) & 2 h \\\hline
\hline
\end{tabular}
\end{center}
\end{table}

\begin{figure}[t]
\centering
\includegraphics[scale=0.50]{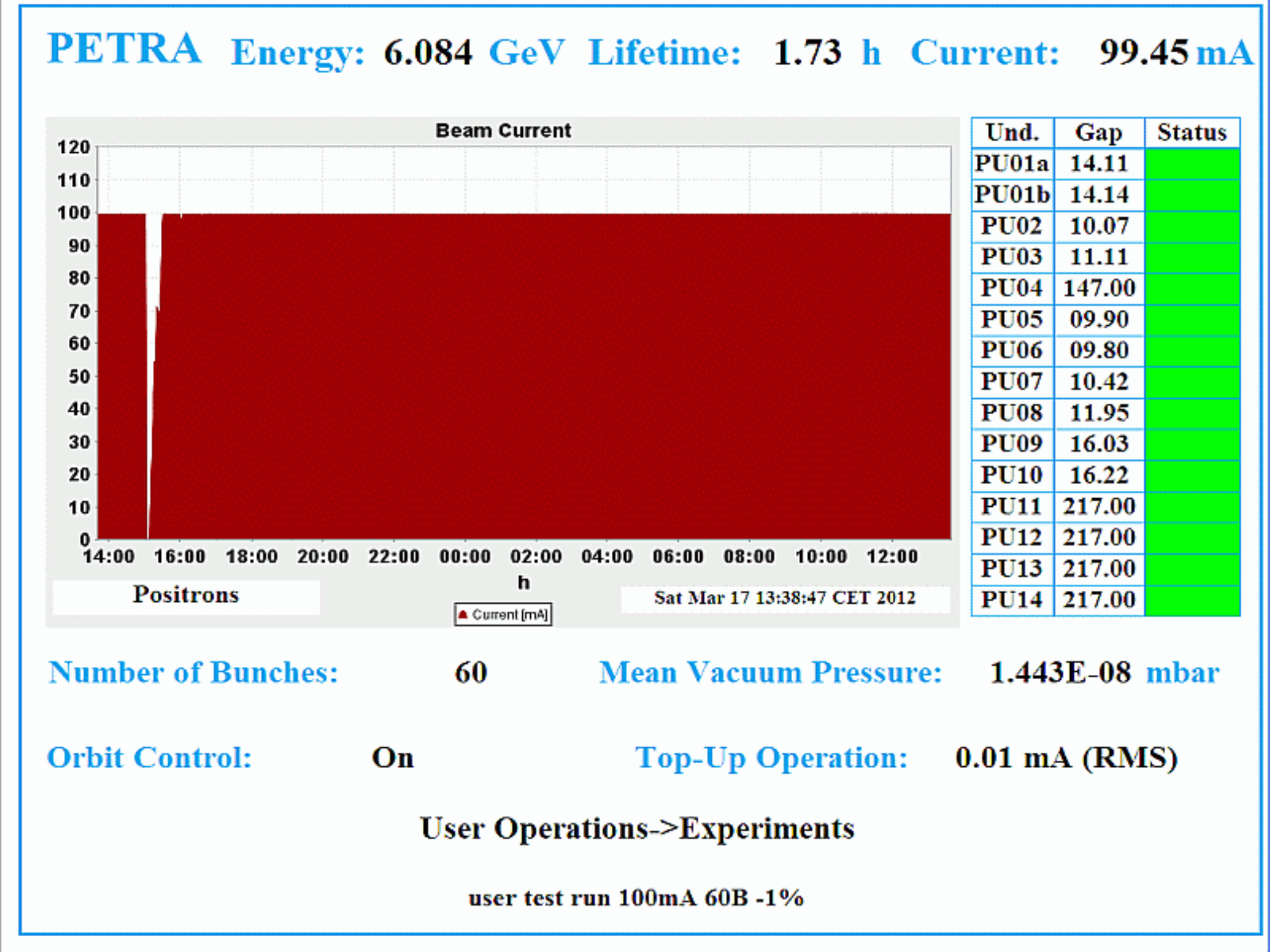}
\caption{\label{online}
PETRA online status dysplay~\cite{p3online}.}
\end{figure}

Time resolved state with 40 or 60 bunches is the main working mode of the machine.
Top-up running allows long-term stable operation with constant current. On  
Figure~\ref{online} a typical performance of the PETRA is shown over a 24 hours period.  

At the planned laser-positron interaction point PETRA beam 
has a  horizontal dispersion $D_x=0.139 m$ and following Twiss parameters \\
    $\alpha_x = 0.427$, $\beta_x = 11.114 m$\\
      $\alpha_y=-1.311$, $\beta_y=19.945 m $ \\
with RMS beam sizes  $\sigma_x =106 \mu m$ and $\sigma_y = 24 \mu m$ \\
and RMS divergences  $\sigma_{x'} =10 \mu rad$ and $\sigma_{ y'} = 2 \mu rad$. 

\subsection {Laser}

Laser in this application should provide sufficient luminosity for high Compton 
rates at single or few particles operational mode.
In addition the light wavelength should be possibly short to extract and detect
the scattered positrons within available limited distance from interaction point.
These two requirements somewhat contradict each other since the Compton 
cross-section is falling toward shorter wavelengths. Thus, at acceptable
wavelengths the laser power demand is so high that appropriate CW lasers are 
not available in market.
Since high power Q-switched lasers are not adequate for single particle mode
applications,  we have to choose among mode-locked lasers.
An example of commercially available laser which meets our needs is Coherent Paladin 
UV laser~\cite{plaser} with parameters listed in the table~\ref{tab2}. 

\begin{table}[ht]
\caption{Coherent Paladin laser specifications.}
\label{tab2}
\begin{center}
\begin{tabular}{|l|l||}
\hline \hline
\multicolumn{2}{|c||}{ \includegraphics[scale=1.5]{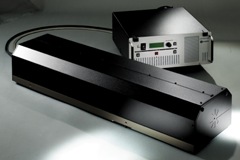} } \\
\hline
Wavelength & 355 nm \\ \hline
Output Power & $> 8$ W\\\hline
Repetition Rate  & $80 \pm 1$ MHz \\ \hline
Pulse Length  & $>15$ ps at 1064 nm\\ \hline
Spatial Mode  & TEM00\\\hline
M2  & $< 1.2$ \\\hline
Beam Diameter  & $ 1 \pm 15\% $ mm \\\hline
Beam Divergence  &  $<550 \mu rad$ \\ \hline
Beam Ellipticity  & 0.9 to 1.1 \\\hline
Pointing Stability   & $<20 \mu rad/ ^\circ C$  \\ \hline
Polarization linear  & $>100$:1, vertical\\\hline
Long-term Power Stability  & $< \pm 2\% $\\ \hline

\hline
\end{tabular}
\end{center}
\end{table}

The laser light will be delivered to the interaction chamber by a  
single mode, polarization maintaining fiber. This will provide fixed position 
of the light at the interaction point independent of the polarization state.
At the end of the fiber a quarter wave plate will convert linear light into
a circular one which then will be focused $<10\mu m$ to the positron beam.
Polarization state of the interacting laser photons will be controlled
by an electro-optical Pockels-cell device installed upstream of the fiber.  
Light polarization and intensity will be constantly monitored at 
laser beam dump, in Analyzer Box. 

\subsection {Experimental sensitivity}

\begin{figure}[!t]
\centering
\includegraphics[scale=0.60]{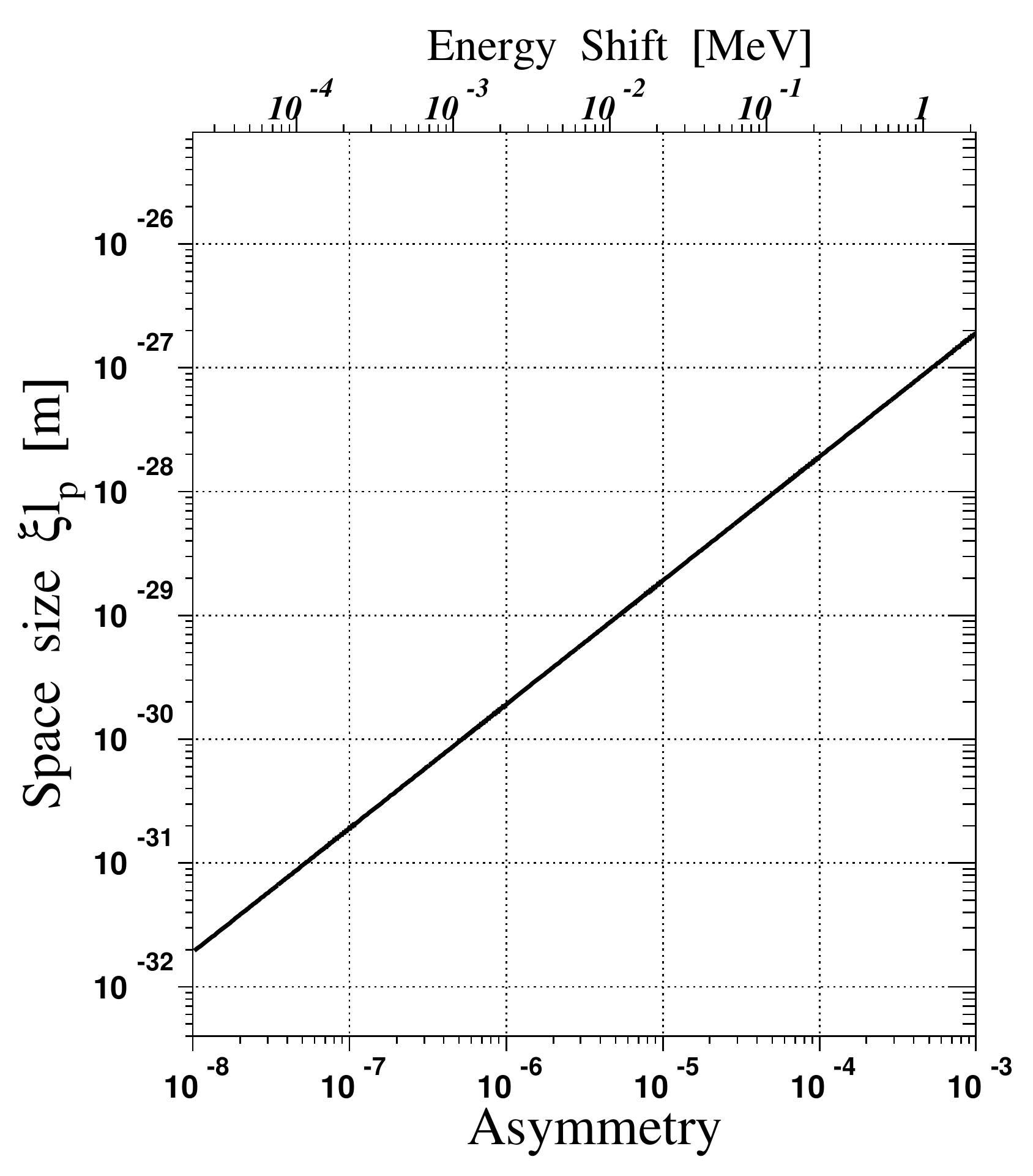}
\caption{\label{fig2}
Sensitivity of the PETRA III for vacuum birefringence and refractivity.
Birefringence at the scale $\xi$ will produce a Compton edge asymmetry (lower scale)
while the refractivity produces absolute energy shifts (upper scale).}
\end{figure}
Given the accelerator energy of 6~GeV, laser wavelength of 355~nm and 
$e\gamma$ crossing angle of $90^\circ$, gravity induced effects could be calculated 
using Eq.(\ref{eq9}) and Eq.(\ref{eq11}). Single refraction in crystal space 
will shift the Compton edge while laser helicity flip will produce an energy 
asymmetry induced by double refraction. Magnitude of these effects are shown 
in Figure \ref{fig2}. Experimental reach of the experiment is then defined
by accuracies for the energy and asymmetry measurements as well as limiting 
systematic effects. We expect precisions of 
 $\Delta \omega_m /  \omega_m = 10^{-3}$ to $10^{-4}$ for energy and 
 $ 10^{-7}$ to $10^{-8} $ for  asymmetry measurements which are corresponding 
to upper right and lower left regions on the  Figure \ref{fig2}.
Detailed calculations will be presented in the folowing sections.  
   
\subsection {Beamline}

\begin{figure}[t]
\centering
\includegraphics[scale=0.52]{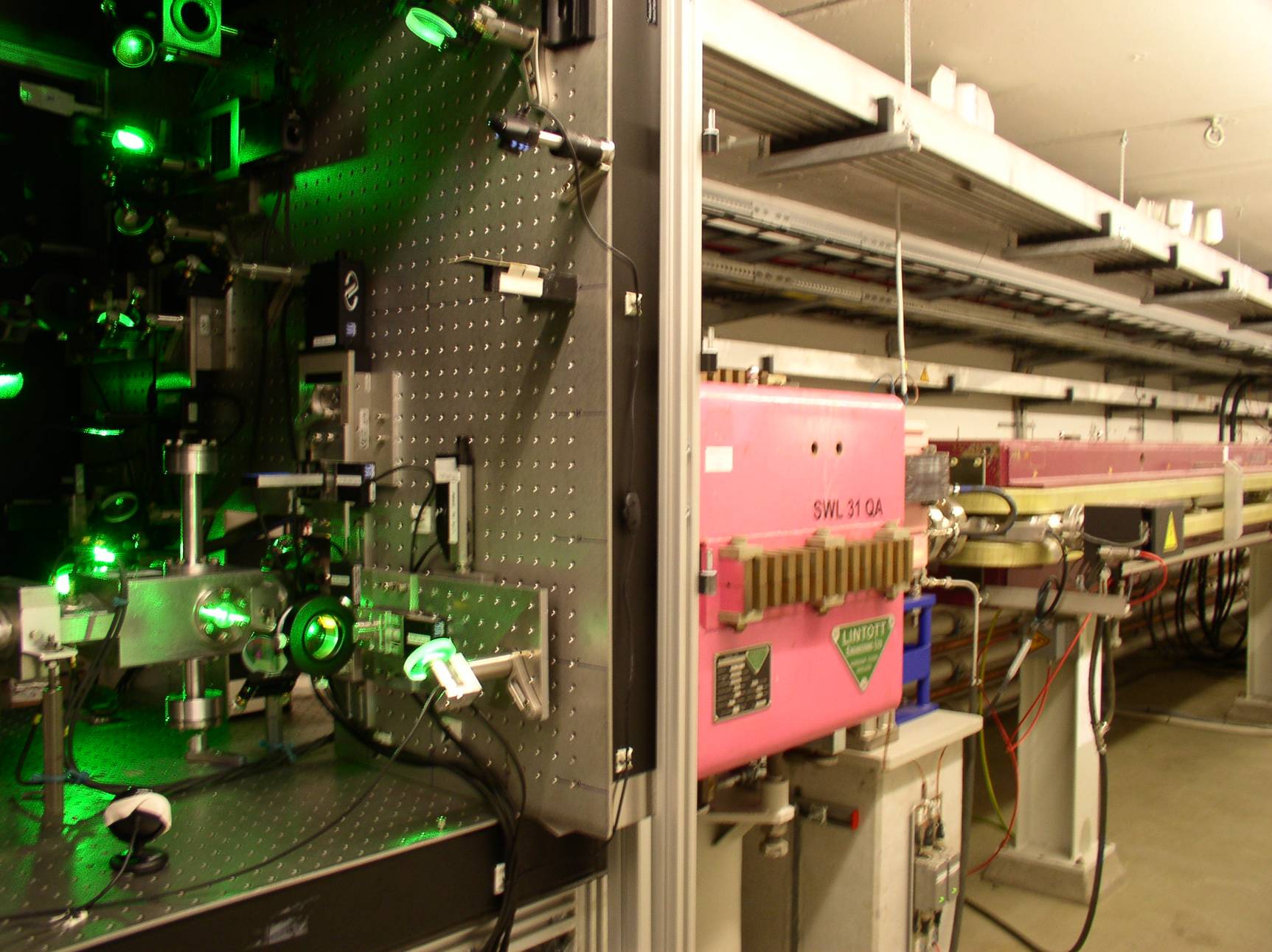}
\caption{\label{bline-photo}
Existing beamline: interaction chamber surrounded by Laser-Wire optics with following quadrupole and dipole magnets. A green laser is shining through
horizontal view port of the chamber.}
\end{figure}

We plan to use existing beamline and  interaction vacuum chamber build for 
PETRA-III Laser-Wire project~\cite{las-wire} (see Figure~\ref{bline-photo}). 
The chamber has horizontal and vertical optical entry-exit ports (windows) 
for the laser.
For our application we will use vertical ports which assumes careful stress-less 
mounting of the laser windows to  preserve circular polarization of the light.

\begin{figure}[!ht]
\centering
\includegraphics[scale=0.80]{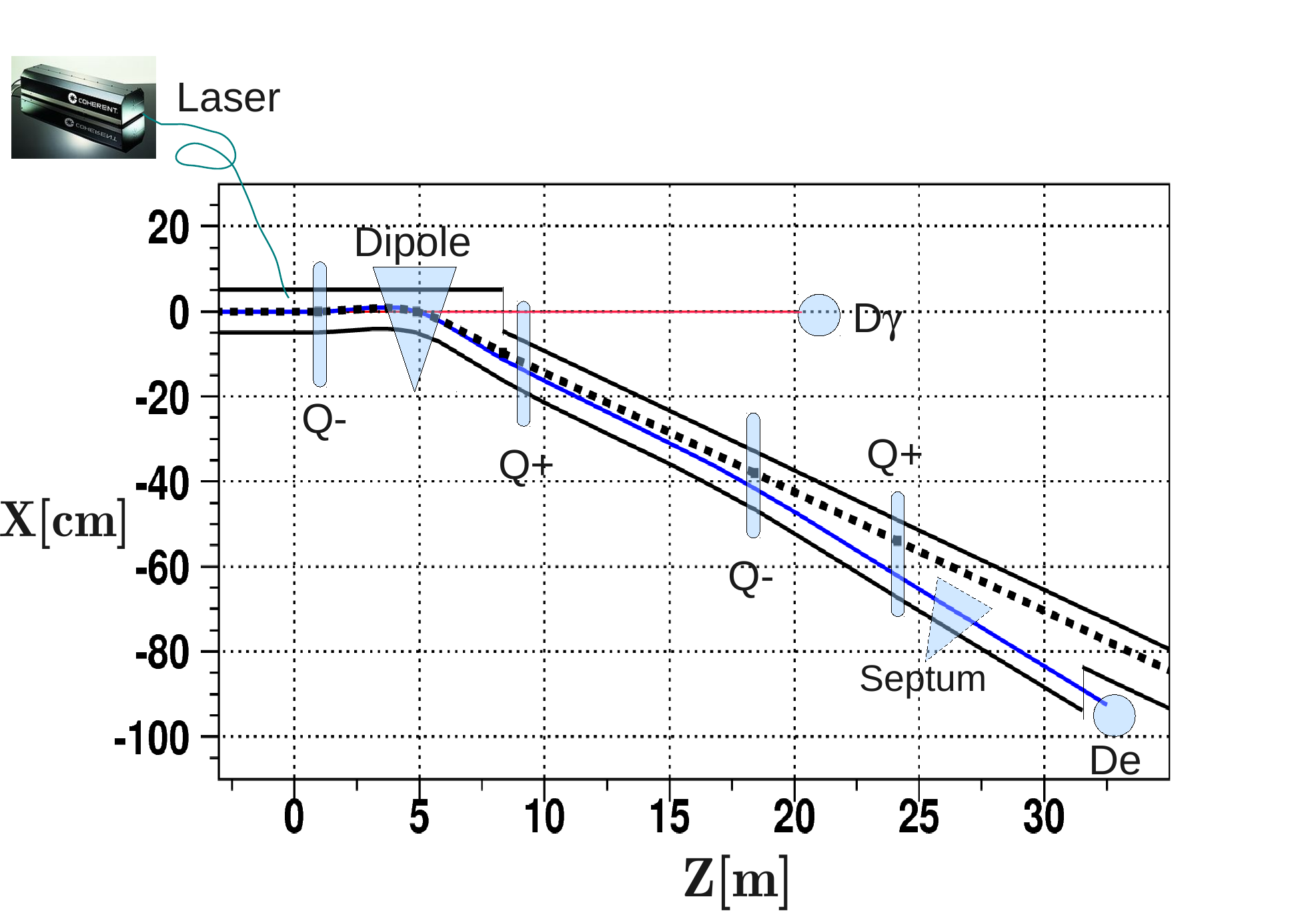}
\caption{\label{bline}
Beamline elements relative to laser and positron beam (dotted line) 
interaction point ($Z=0m$, $X=0cm$). Beam pipe and exit windows are drawn in black,
trajectories for Compton edge positrons and photons are shown in blue and red.
Q(-)+ are assigned to (de)focusing  quadrupoles. Photon and electron detectors 
denoted by $D\gamma$ and $De$ respectively.}
\end{figure}

\begin{table}[ht]
\caption{Positions of beamline components.}
\label{tab3}
\begin{center}
\begin{tabular}{|l|c|l|l||c|c||}
\hline
\hline
 Component  & DB   & Position & Position &$\gamma - e$ beam  &  $e^{\prime}-e$ beam \\
  name      & name &  SWL(m) & vs IP(m)  &separation & separation \\\cline{1-4}
   BPM-SL1 & BPM & 38.8295 & -6.5055 & $X_B-X_\gamma$ &  $ X_e - X_B$ \\ \cline{1-4}
   BPM-SL2 & BPM & 32.6000 & -0.2760 & (cm) & (cm) \\\hline\hline
   IP & LSW & 32.3240 & 0.0000 & 0.0000 & 0.0000 \\\hline
   Q- & Q5K & 31.3670 & 0.9570 & 0.0000 & 0.0000 \\\hline
   Pickup & BPM & 30.9365 & 1.3875 & 0.0000 & 0.0000 \\\hline
   Dipole & DK & 27.4410 & 4.8830 & 4.8288 & 0.7745 \\\hline
   Q+ & Q4K & 23.9980 & 8.3260 & 9.6576 & 1.5489 \\\hline
   Q- & Q2K & 13.9000 & 18.4240 & 37.9824 & 3.6515 \\\hline
   Q+ & Q1K & 8.2000 & 24.1240 & 53.9708 & 8.2569 \\\hline  
   BPM-$X_B$ & BPM & 0.6000 & 31.724 &  &  \\\hline
 e-window & - & 0.8240 & 31.8500 & 74.6604 & 14.2165 \\
\hline

\end{tabular}
\end{center}
\end{table}

A vacuum exit  window (2mm Al) for the Compton photons is located at 7.8m
distance from the interaction chamber.    
Beamline essential components are listed in table~\ref{tab3} and are drawn in 
Figure~\ref{bline}. 

Beam position monitors (BPM) near and around the interaction point 
are dedicated for beam position, slope (BPM-SL1, BPM-SL2) and bunch timing (Pickup) 
measurements~\cite{Chrin:2009zz}.  
Beam horizontal position evaluated by BPM-$X_B$ at SWL 0.6m will enter space refractivity 
derivations.

Most important beamline element, apart the interaction chamber, is a 5.4m dipole
which will separate scattered positrons and photons from the PETRA (neutral) beam. 
Focusing and defocusing quadrupoles are assigned by $Q+$ and $Q-$ respectively,
relative to the horizontal (x) plane. Quadrupoles located  downstream of the dipole  
will noticeably bend the Compton edge positrons because of considerable horizontal 
offsets from the quadrupole center. Such a bend is visible at the last defocusing 
quadrupole in the Figure~\ref{bline}. Last focusing quadrupole bend should be compensated
to achieve necessary separation between the extracted positron and neutral beam at 
the exit window. 
 Characteristics of the dipole and quadrupole magnets~\cite{p3param} are displayed in
table \ref{tab4} for the 6~GeV machine and nominal beam optics conditions.   

Separations between neutral beam and scattered photons and positrons  
are also included in the table~\ref{tab3} (last 2 columns). These distances are
calculated at the geometrical centers of the magnets along z. 
Table shows sufficient outside (the ring) room to place a $\gamma$ detector
starting about 25m downstream of the interaction point
while the Compton positrons could be comfortably detected at 31.5m, inside
the ring.  

\begin{table}
\caption{Magnet parameters.}
\label{tab4}
\begin{center}
\begin{tabular}{|l|l|}
\hline
\multicolumn{2}{|c|}{Dipole} \\
\hline
Length & 5.378 m \\

Bending angle & $1.607^\circ$ \\
Bending radius & 191.73 m \\
Field & 0.10439 T \\
Field Error $\Delta B/B$ & $ 5 \cdot 10^{-4}$\\
Critical Energy & 2.499 keV \\
\hline
\multicolumn{2}{|c|}{Quadrupoles} \\
\hline
Length & 1042 mm \\
Aperture & 50 mm\\
Gradient (max) & 15 T/m\\
k & $0.749 m^{-2}$\\
Field Error  $\Delta k/k$ & $4\cdot 10^{-3}$\\
\hline
\end{tabular}
\end{center}
\end{table}

Vacuum beam pipe downstream of the Q2K  should be modified to allow extraction 
and detection of the Compton scattered positrons. For that the vacuum pipe has 
to be extended on inner side as it seen on the Figure~\ref{bline}.
The extension will end by a Titanium exit window of thickness $356 \mu m$ 
and size $60mm\times 20mm$. In addition magnetic field of the last quadrupole
Q1K should be shielded for scattered positrons. Otherwise focusing field of the 
quadrupole is considerably strong for off-center Compton edge positrons
to bring them back to the  beam. 
In case of technical difficulties for quadrupole field shielding, it is possible 
to use an additional dipole magnet.
Right after the Q1K quadrupole there is sufficient separation (about 7~cm) between 
the Compton positrons  and neutral beam to accommodate a septum magnet.

In contrast to space birefringence, experiment for the refractivity  requires
absolute position measurements of the scattered Compton particles. Therefor, 
positioning and alignment of the 
$\gamma$, $e^+$ detectors and the laser beam should be done with best available 
accuracy in horizontal plane. For absolute calibration of the BPM-XB a horizontal
wire scanner should be installed in near vicinity of the BPM.

\section {Expected performance}

\begin{figure}[t]
\centering
\includegraphics[scale=0.75]{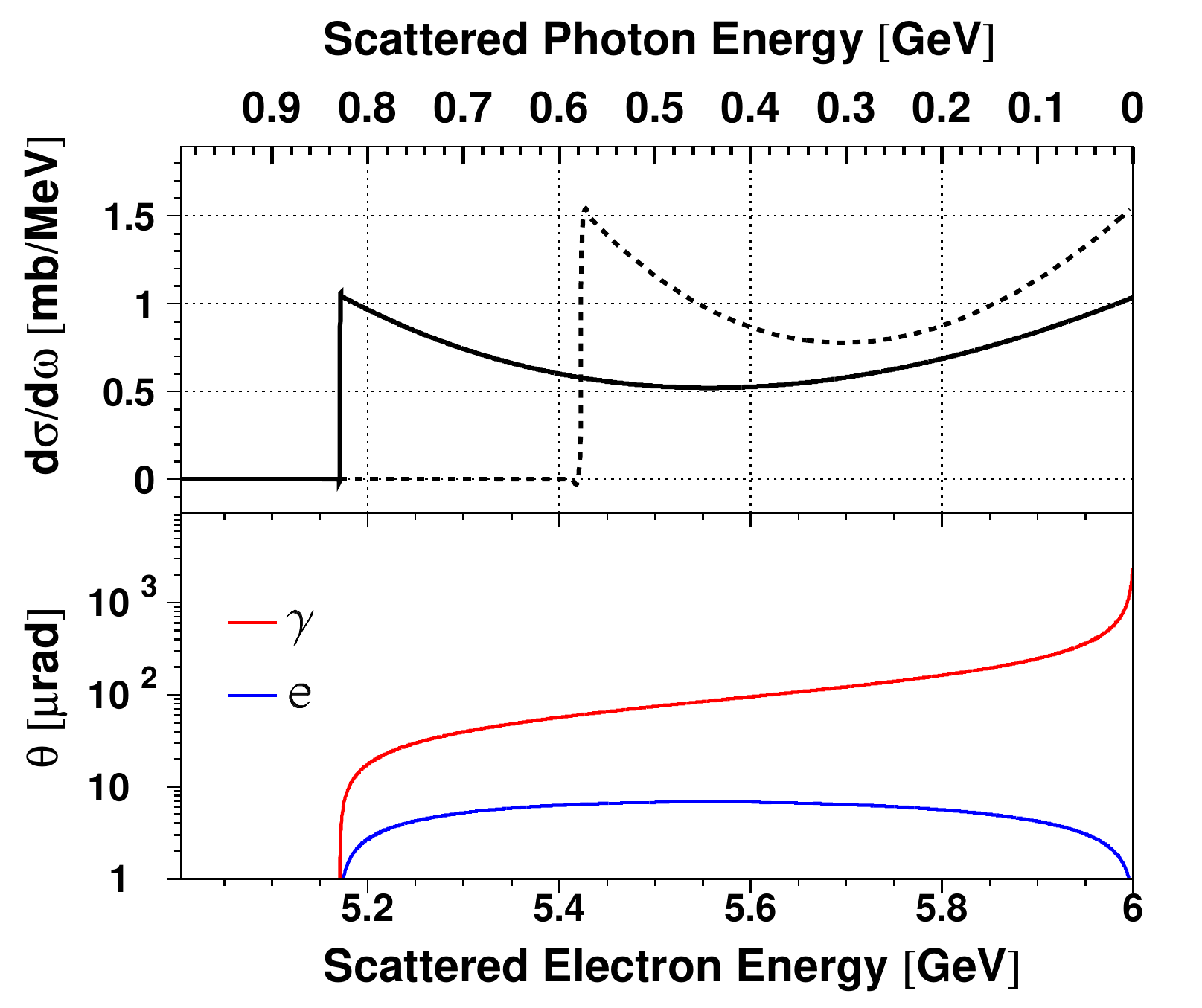}
\caption{\label{spec1}
Compton cross-section and scattering angles for initial 6 GeV positron and 
3.49~eV photon. Laser and $e^+$ beam crossing angle is $90^\circ$. 
Dotted line spectrum is for 2.33~eV green laser.}
\end{figure}
\subsection {Compton spectra}

For 6~GeV PETRA and 355~nm laser, energy and angular distributions 
of the Compton positrons and photons are presented on Figure~\ref{spec1}. 
Scattering kinematic factor is x=0.16 which corresponds to Compton edge positron
(minimal) energy of $\mathcal E'_{min} = 5.17~GeV$.  
At this energy the scattered positron retains initial movement direction  
as it is clear from the plot displaying angular dependencies.   
Compton edge photons follow the same direction with an energy of 0.83~GeV. 

Hunted gravitational effects will change energy sharing between the positron 
and photon. Expected changes
are relatively small and will be hardly detectable by direct calorimetry so,
we will use beamline magnets to convert scattered positron energy (momentum) 
to position in order to explore more sensitive instrumentation.
To estimate  spatial relationships we apply an approximate formula
connecting energy and position of the scattered positron.  
Detected horizontal position of the scattered positron with energy ${\mathcal E'}$ 
could be presented by 
\begin{equation}
X= X_0 + (Z-Z_D) \left(\theta_x+ \frac{eLB}{{\mathcal E'}}  \right) 
\label{horcor}
\end{equation}
where $X_0$, $\theta_x$ are position and horizontal angle of the positron 
at the laser interaction point, $Z$ and $Z_D$ are locations of the detector and
bending dipole respectively. Here $L$ and $B$ stand for the 
dipole length and magnetic field while influence of the quadrupoles is ignored.
From this relation it follows that an energy change $\Delta \mathcal E'$ around the 
Compton edge will produce a position change 
$$\Delta X = 144.3 \Delta \mathcal E' \mu m/MeV$$
at the detector location Z=32m.

\subsection {Smearing factors}

Momentums and energies of Compton particles are smeared by initial laser
and positron beam position, angular and energy distributions.
We use Eq.\ref{horcor} with $e^+$ beam optics parameters and numbers from Table \ref{tab1},
Table \ref{tab2} to calculate magnitudes of smearing factors. 
Estimated influence of different factors on position of the Compton edge positron
at detector location is shown in Table \ref{tab5}. 

\begin{table}[h]
\caption{Smearing factors.}
\label{tab5}
\begin{center}
\begin{tabular}{||l|l|c||}
\hline
 Factor & Value & $\mathcal E'$ smearing  \\
%\multicolumn{2}{|c|}{Dipole} \\
\hline
 $e^+$-laser IP & $\sigma_{IPx}= 10 \mu m$ &   $10 \mu m$ \\
 $e^+$ energy spread & $\sigma_{\mathcal E}/{\mathcal E}= 10^{-3}$ &   $750 \mu m$ \\
 $e^+$ divergence & $\sigma_{x'}=10 \mu rad$  & $280 \mu m $ \\
e-window & $ \sigma_{Mult}=2.4 mrad$ & $50 \mu m  $  \\
\hline
\end{tabular}
\end{center}
\end{table}

\begin{figure}[!ht]
\centering
\includegraphics[scale=0.70]{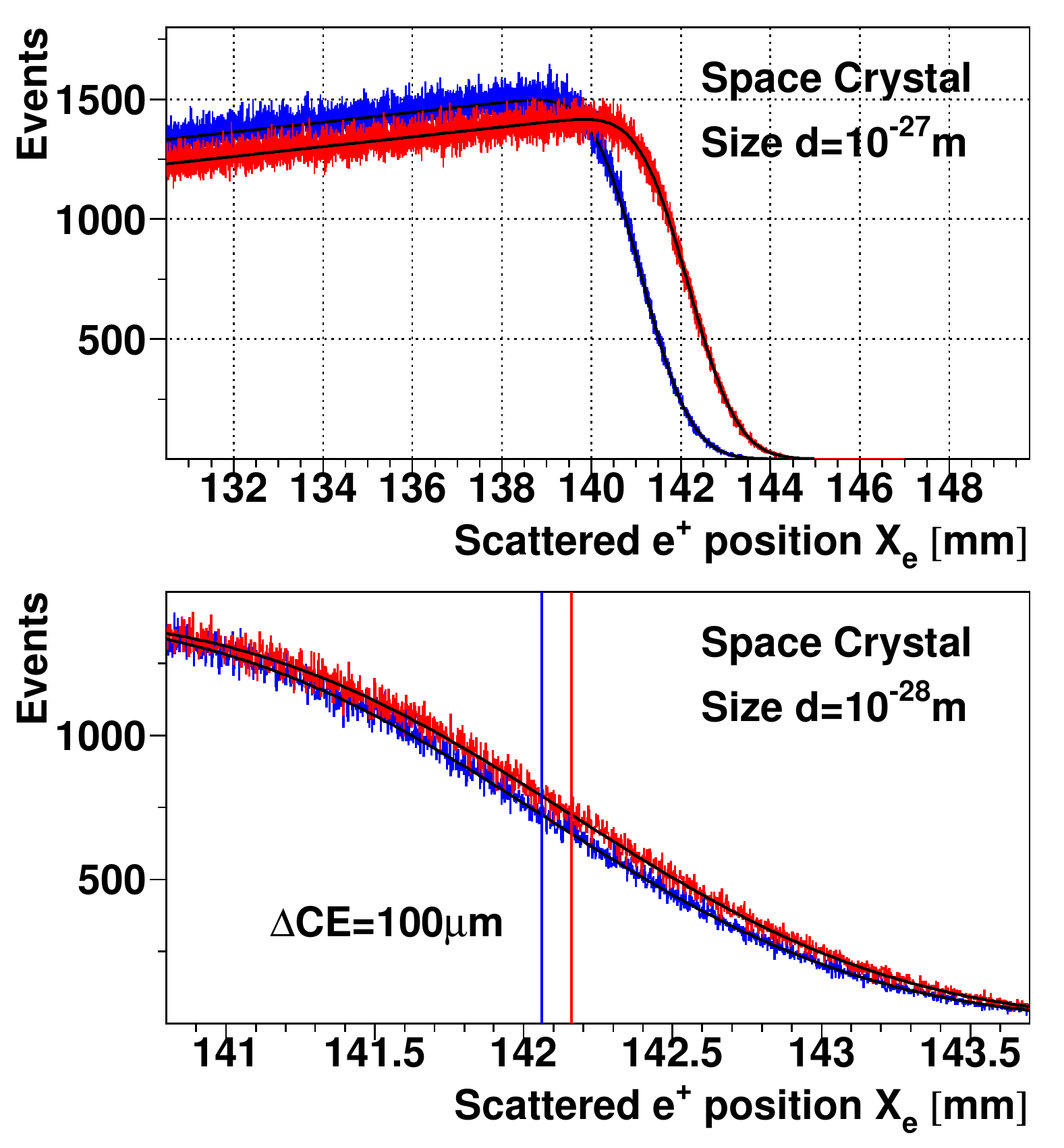}
\caption{\label{espectr}
Simulation results: scattered positron spatial spectra for laser left and right helicities. Horizontal scale is distance between the scattered $e^+$ and neutral beam
at the detector location. Lower plot shows Compton Edge (CE) positions obtained 
by fitting a 4 parameter function from ref.~\cite{Klein:1997wq} to spectra. Same
function is applied for fitting upper plot distributions.} 
\end{figure}

Apart from the mentioned main contributing factors Table \ref{tab5} displays also smearing 
by multiple scattering in the positron exit window.
Overall smearing is about $800 \mu m$ which agrees to a detailed simulation results
presented in Figure \ref{espectr}. Dominant smearing contributor is the lepton beam 
energy spread, quantifying as
 $\sigma_{x{\mathcal E}} \approx \Delta X_{ee'} \sigma_{\mathcal E}/{\mathcal E} $
with $\Delta X_{ee'}$ being separation between scattered positron and neutral beam at 
detector location. 
Hence, one can reduce smearing induced by inter-bunch energy spread only via moving 
detector closer to IP with an expense of shrinking available place for the $e'$ detectors. 

%divergence which is a result of relatively small $\beta$ at the interaction point.
%Other, high $\beta$  PETRA regions for the proposed experiment, may considerably 
%improve situation - for instance, at NWR 23m smearing will be reduced by a factor of 4.

\subsection {Rates}

Compton secondaries will be detected in single particle resolving regime
with about 0.01 particles per bunch. 
PETRA mostly operates with 40, 60 or 240 bunches~\cite{acc2011} with corresponding
inter-bunch spacings of 192ns, 128ns and 32ns.   
Therefor, expected Compton rates are 52kHz, 78kHz and 313kHz for different 
bunch modes of the machine. However, since we are interested exceptionally on 
the Compton edge particles, the rates can further be reduced by discriminating
energies of the photons or positrons. Assuming 5\% energy detection resolution
for single particles the rates could be reduced by a factor of 3 or 12 triggering 
on the positron or photon calorimeter respectively.  This numbers are derived 
by integrating spectrum on the Figure \ref{spec1} within ranges of 0 to 100\% 
and 95\% to 100\%.
 
\subsection {Detectors}

\begin{figure}[!ht]
\centering
\includegraphics[scale=0.70]{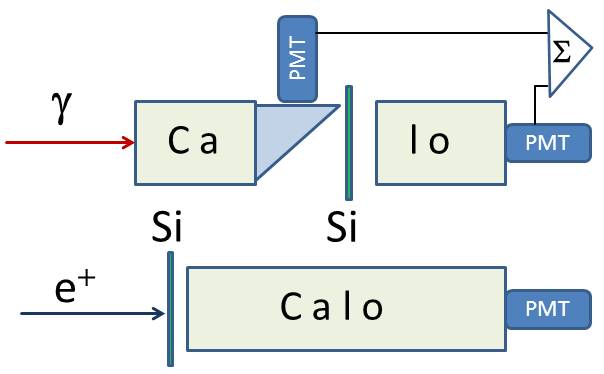}
\caption{\label{detect}
Photon and positron detectors schematics. Position sensitive part is denoted by
"Si" and energy sensitive part by "Calo". PMT stands for photomultiplier.} 
\end{figure}

In previous sections we have defined energy and rate of the expected signals.
For $e^+$ and $\gamma$ produced at Compton edge, simultaneous position and energy 
measurements are necessary. We intend to install a combination of position sensitive 
and calorimetric detectors at the positron and photon branch. 
For energy measurements homogeneous crystal calorimeters could provide a resolution
of 5\% over $\sqrt{{\mathcal E}}$. Position measurements will be performed 
with silicon strip or pixel detectors. A position resolution of $10 \mu m$ and 
rate capability of 30 kHz would be sufficient for the whole range of our measurements.
The silicon detector will be placed in front of calorimeter for positrons 
while for the photons position detector will be located at middle part of the 
calorimeter where shower lateral size is maximal. This arrangement is displayed 
in Figure \ref{detect}.
   
\subsection {Backgrounds}

Apart from the laser Compton scattering there are other beam related sources of
scattered photons or leptons at any accelerator environment. These particles may
enter detectors and spoil measured distributions. For storage rings one should 
first consider synchrotron radiation from bending or quadrupole 
magnets~\cite{Chao:1999qt}. 
In our case positron detectors are located inside the ring and could see only scattered
synchrotron light while the ouside photon detectors are imposed to direct synchrotron 
radiation. Therefor we plan to shield detectors at the beam pipe side and, 
in addition, the photon detectors also at the front side to absorb completely 
the synchrotron radiation.       

Other potentially dangerous
processes are beam-gas interaction~\cite{LeDuff:1986pa}, 
scattered blackbody radiation~\cite{Telnov:1986ki,Dehning:1990tb} 
and intra-bunch scattered (Touschek) positrons~\cite{Bernardini:1997sc}.
Explored positron and photon coincidence registration mode will greatly suppress
or completely eliminate backgrounds from the thermal photons and Touschek positrons.
The beam-gas Bremsstrahlung, however, can not be discriminated since the energy
balance is similar to the Compton scattering. This would be the main background 
and it should be handled by keeping vacuum pressure around the Compton interaction 
point possibly low (mounting of additional vacuum pumps should be foreseen). 
The Bremsstrahlung rate will be monitored periodically by blocking the
laser light by a shutter.  Alternatively a fast, electro-optical or acousto-optical 
modulator may be used to redirect the laser out of 
certain portion of bunches for background measurements. 

A more severe source of background at the SW section  could be  
aperture limitations - beam collimators (coll1 and coll2) of PETRA are located about 15m
upstream and downstream of our detectors. 

\subsection {Measurements}

\subsubsection {Space birefringence }

As it was described above for space chirality measurements the laser 
helicity will be flipped  with few hundred Hertz sweeping frequency to 
avoid correlations with any possible periodic source. 
Accumulated spatial events will be tagged with the helicity and the
resulted spectra (simulated samples are shown on Figure~\ref{espectr}) 
will be analyzed to fetch the Compton edge for each helicity.
For that the spectra could be fitted by  
gaussian and error combined functions as it is 
proposed in~\cite{Klein:1997wq} or by Compton cross-section convoluted 
with detector resolution, as in~\cite{Gharibyan:2003fe}. 
An example of fitted spectra is shown in  Figure~\ref{espectr}, where
a Compton edge shift of $100\mu m$ is detected by fits for simulated initial
space birefringence at $\xi = 10^7~(\approx 10^{-28}m)$ .
After the fits an asymmetry 
\begin{equation}
A=\frac{{X_e}^+ - {X_e}^-}{{X_e}^+ + {X_e}^-}
\label{eq13}
\end{equation}
will be calculated
with ${X_e}^+$, ${X_e}^-$ being Compton edge positions for positive and negative 
helicities. Finally, this measured asymmetry will be related to Eq.(\ref{eq11}).

\subsubsection {Space refractivity }

For vacuum index measurement it is necessary to detect absolute energy
of the Compton edge positron or photon. 
 This could be accomplished by
simultaneous  measurements of the photon $X_{\gamma}$, positron
$X_e$, and neutral beam $X_B$ positions 
\footnote{Method is similar to a 3-positions measurement scheme proposed 
for ILC energy determination~\cite{Muchnoi:2008bx}. }
(see Figure~\ref{gscheme}). 
With this information one can use Eq.(\ref{eq8}) and Eq.(\ref{horcor}) 
to arrive to
\begin{equation}
n-1=\frac{1}{2 \gamma^2} \left( x\frac{(X_B - X_{\gamma})^2}{(X_e-X_B)(X_e-X_{\gamma})} -1  \right) 
\label{birefex}
\end{equation}
which holds if the scattered positron and the neutral beam are transported
through the same (strength) magnetic field, i.e. for homogeneous dipole field.
With quadrupoles one needs to apply corrections which could be measured
during calibrating quadrupole scans.
This is, still, not the full story, since
any small offset in beam energy value ${\mathcal E}_{beam}$
would result in a fake refractivity measurement. Therefor, for space refractivity experiment it is necessary an independent, precise measurement of the beam absolute energy. Moreover, since the beam energy changes along the ring, the energy 
measurement should be done at the Compton interaction point. For this we will 
explore a different frequency light generated by the same or another laser. 
Since the UV 355nm light is third harmonic of the 1064nm Nd:YAG laser, we can use
second harmonic, 532nm wavelength which otherwise is widely available as a standalone
laser solution.
 
Combining Eq.(\ref{eq8}),(\ref{eq7}) and Eq.(\ref{horcor}) for two laser photon energies, after some lengthy though simple calculations we obtain expressions for 
beam energy and refractive space size measurements 
\begin{equation}
  {\mathcal E}_{beam}\left[ GeV \right] = \frac{  u_1 +u_2 L_2 + u_3 L_1 +u_4 L_1 L_2}
    {u_5 + u_6 L_2 + u_7 L_1 + u_8 L_1 L_2}
\label{eqen}
\end{equation}
\begin{equation}
\zeta l_P \left[ m \right]  = 7 \cdot 10^{-24}
\frac{v_1 + v_2 L_2 + v_3 L_1 + v_4 L_1 L_2 + v_5 L_1^2  +v_6 L_1^2  L_2}
{v_7 + v_8 L_2 + v_9 L_1 + v_{10} L_1 L_2+v_{11} L_1^2  +v_{12} L_1^2  L_2}
\label{eqref}
\end{equation}
where $L_1, L_2$ are incorporating position measurements for the laser 
1 (UV, 3.49~eV) and laser 2 (green, 2.33~eV) and
\begin{equation}
 L = \frac{ X_e - X_B}{X_B - X_\gamma}
\label{eqll}
\end{equation}
is a common expression to calculate $L_1,L_2$.
Coefficients $u$ and $v$ depend solely on two laser wavelengths, crossing
angles and a central energy of $e^+$ beam chosen to be ${\mathcal E}_0 = 6.00000~GeV$. 
Resulting expressions are too long to be presented here, hence we display
numerical values of the coefficients in Table \ref{coeff2}. 

\begin{table}[h]
\caption{Coefficients in Eq.(\ref{eqen}) and Eq.(\ref{eqref}).}
\label{coeff2}
\begin{center}
\begin{tabular}{||l|l||}
\hline
$u_1=  0.03747513690$ &$u_2=-8.245513915 $ \\ \hline
$u_3= 3.281272436 $&$u_4= -5.001716619$ \\ \hline
  $u_5= -0.04417775880$ &$u_6= 0.8269285477$ \\ \hline
$u_7= -0.7135896078 $&$u_8= 0.1575166987$ \\ \hline\hline
 $ v_1 = 0.00100246550  $ & $ v_2 = -0.2205687272 $ \\ \hline 
 $ v_3 = 0.1319522859  $ & $ v_4 = -0.9607252134 $ \\ \hline 
 $ v_5 = 0.7135896078  $ & $ v_6 = -0.1575166987 $ \\ \hline 
 $ v_7 = -0.04417775880  $ & $ v_8 = 0.8269285477 $ \\ \hline 
 $ v_9 = -0.7577673666  $ & $ v_{10} = 0.9844452464 $ \\ \hline 
 $ v_{11} = -0.7135896078  $ & $ v_{12} = 0.1575166987 $ \\ \hline 
\end{tabular}
\end{center}
\end{table}

Described formalism allows to measure space refractivity provided 
UV and green lasers to be delivered to the same interaction point.
Spatial separation of Compton edge positrons from two lasers will be 
40.915~mm thus allowing to use the same exit window and $e^+$ detector.    

\subsubsection {Space anisotropy }

Possible spatial dependence of space birefringence or refractivity  
will be tested by writing Eq.(\ref{eq12}) for PETRA declination angle
$\delta=22.58^\circ$ at interaction point 
\begin{equation}
Q = Q_0 ( 0.92 \cos{\delta_0}\cos{(\alpha-\alpha_0)}+0.38 \sin{\delta_0})
\label{eqiso}
\end{equation}
with $Q=\Delta n$ for measured birefringence or $Q=n-1$  for refractivity. 
For each measurement time a corresponding right ascension angle $\alpha$
will be calculated and obtained $Q-\alpha$  dependence will be fitted by
Eq.(\ref{eqiso}) to find space anisotropy axis direction $\delta_0,\alpha_0$.

\section {Experimental reach and accuracy }

\subsection {Statistical errors }

For an expected small asymmetry from Eq.(\ref{eq13}) 
${X_e}^+ \approx {X_e}^-  = X_e \approx 142mm$, error propagation gives
\begin{equation}
\Delta A = \frac{1}{\sqrt{2}}\frac{\Delta X_e}{X_e}
\label{eq15}
\end{equation}
where $\Delta X_e$ is accuracy of the Compton edge ($X_e$) measurement.
Although  Compton edge is derived by fitting distribution with many events,
for statistical error estimation it is more convenient evaluating a single event
accuracy which allows a direct application of conventional statistical events-strength
formalism.  Thus, we can assign
$\sigma_{X_e}\approx 800 \mu m$ as position error for a single event  
equal to the position smearing derived above, and get 
$\Delta X_e=\sigma_{X_e}/\sqrt{TR_e}$, where $R_e$ is rate of $e^+$ events
around  Compton edge and $T$ is time of measurement. 
Necessary data taking times to achieve different sensitivities, for 
an average rate of $R_e=13kHz$ (estimated from 2011 running~\cite{acc2011}),
 are displayed on Table~\ref{astat}.
\begin{table}[h]
\caption{Asymmetry measurement times and space birefringence sensitivities.}
\label{astat}
\begin{center}
\begin{tabular}{||c|c|c||}
\hline
$\Delta A $  & $\xi l_P $  &  T  \\ \hline \hline
$10^{-5} $  & $10^{-28}$m  &  12~sec   \\ \hline
$10^{-6} $  & $10^{-29}$m  &  20~min   \\ \hline
$10^{-7} $  & $10^{-30}$m  &  34~hours   \\ \hline
$10^{-8} $  & $10^{-31}$m  &  141~days   \\ \hline
\end{tabular}
\end{center}
\end{table}

On a way to calculate refractivity measurement errors we estimate spreads 
of Eq.(\ref{eqll}) constituents. A Compton photon position at detector
location, $X_\gamma$, defines initial angle $\theta_x$. Hence, the difference
$X_e-X_\gamma$  will be free from fluctuations of the $\theta_x$.
The beam position $X_B$ is a measure of magnetic field strength
which is completely defined if we explore beam direction (slope) at the
interaction point, measured by two upstream BPMs. From BPM resolution of
$5 \mu m$ per bunch traverse~\cite{Kube:2008zza} we arrive to following 
accuracy estimators for a Compton scattering event:\\
$\sigma (X_\gamma) = 20 \mu m$ \\
$\sigma (X_B) = 15 \mu m$  \\
$\sigma (X_e) = 752 \mu m$ \\
the latter is a quadratic sum of smearings by beam energy and exit window 
from the Table \ref{tab5}. 
Error propagation applied to Eq.(\ref{eqref}), with
derived numbers, yields an accuracy of $5\cdot 10^{-26}m$ for $\zeta l_P$ 
from two (UV and green) Compton edge scattering events.
Evaluating Eq.(\ref{eqen}) with the same events we find  a
$300~MeV$ statistical error for ${\mathcal E}_{beam}$. At PETRA 
this would apply 50~sec data taking time at 5~kHz for beam energy measurement 
with a $10^{-3}$ relative statistical error. A sensitivity of  
$10^{-28}m$ for space refractivity will be achieved during the same time period.

\subsection {Systematic effects}
\subsubsection {Space birefringence }

In general, it is a difficult task to mention asymmetry limiting sources a priori
since most (theoretically all) of beam and detector parameters and their drifts are
not (should not be) affected by helicity flips and are ignorable for asymmetries.    
Therefor, we refer to asymmetry measurement achievements of former accelerator experiments. 
Asymmetries as small as $10^{-7}$ have been detected with a sensitivity $10^{-8}$ 
at the SLAC 50~GeV experiments~\cite{Anthony:2003ub,Anthony:2005pm},   based
on beam helicity flips. 
Same order sensitivities for measured asymmetries are reported at the MAMI 1~GeV 
experiments~\cite{Maas:2005fk,Maas:2008zzc}.
Thus, similar accuracies seem reachable at PETRA  which suggests
that the 6~GeV machine could test  space birefringence down to $10^{-31}m$. 

\begin{figure}[!ht]
\centering
\includegraphics[scale=0.70]{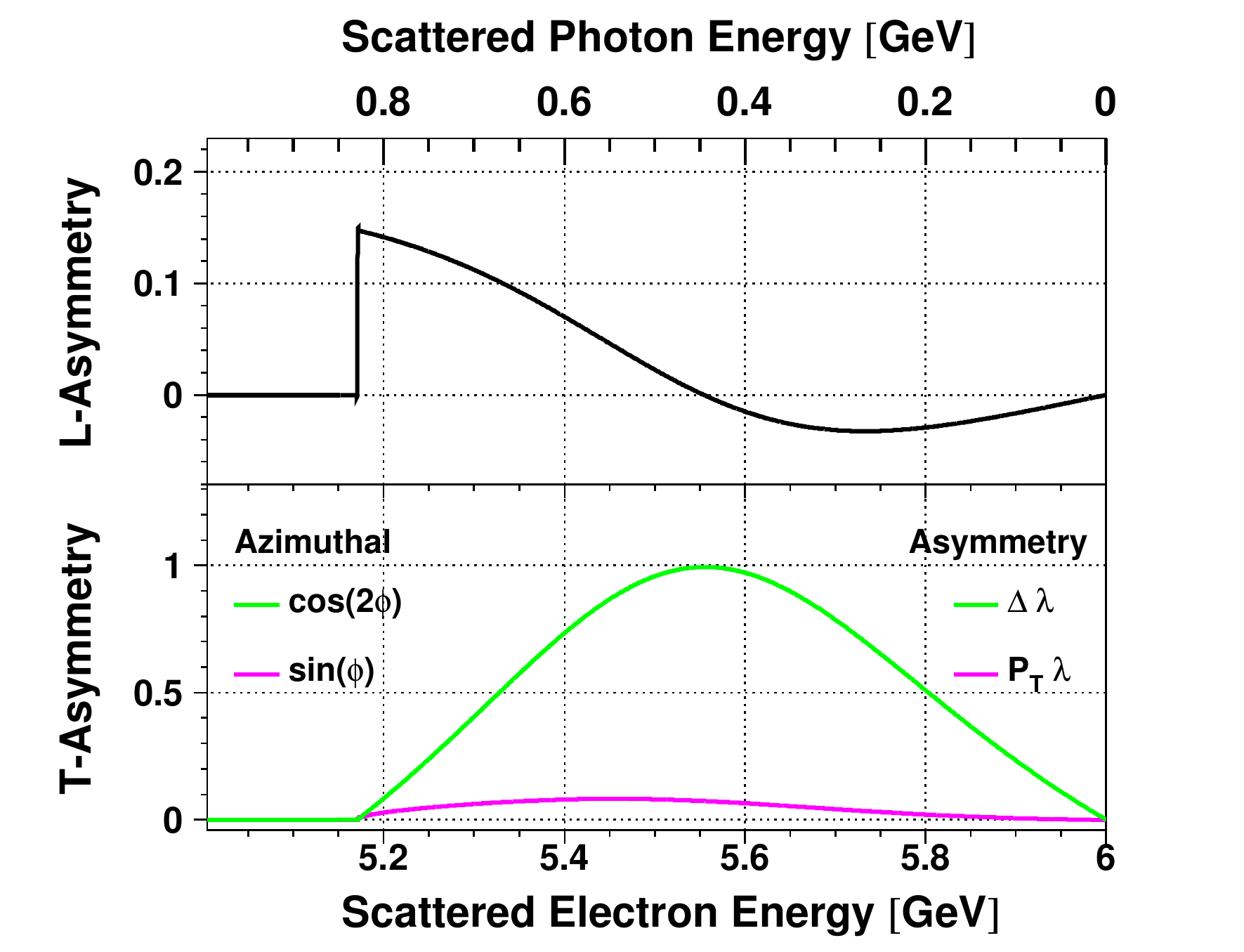}
\caption{\label{asym}
Compton cross-section asymmetry produced by laser helicity flip. Upper plot: for longitudinal positrons. Lower plot: for transversal (magenta) or
unpolarized (green) positrons. In latter case laser polarization is linear. } 
\end{figure}

There are few potential sources of false asymmetry which are 
correlated with laser helicity flips and are all related to
either the laser or lepton beam polarization. These are positron beam 
longitudinal or transverse polarization and laser light linear polarization.
Introduced asymmetries by the mentioned factors, for 100\% polarization, are 
plotted on Figure \ref{asym}. First we note that none of the quoted factors
could shift the Compton edge, although, intensity changes displayed on the 
Figure \ref{asym}, convoluted with detectors responses, could mimic a shift
of the edge.  However, positron beam longitudinal polarization in PETRA should
be plain zero - otherwise the proposed setup is able to measure and monitor even 
small amounts of it. Situation is different for the circular laser light where always 
a small fraction of linear component exist~\cite{Barber:1992fc}
 and for the PETRA beam transverse polarization which could be acquired by 
Sokolov-Ternov mechanism~\cite{Sokolov:1963zn}. 
Nevertheless, since intensity changes produced by helicity flips are vanishing 
at the Compton edge, contribution of these two factors to gravitation induced 
asymmetry should be negligible. 

\subsubsection {Space refractivity } 
Limiting factor for refractivity measurement is positron beam energy 
uncertainty $\Delta {\mathcal E}$. We can calculate corresponding  
effect on refractivity measurement using Eq.(\ref{eq8}) by explicitly writing
$\gamma$ and $x$ dependence on $ {\mathcal E}$ at the Compton edge. Resulting expression
\begin{equation}
\Delta (n-1) = \frac{1}{2\gamma^2}\frac{\Delta {\mathcal E}}{{\mathcal E}}
\label{eq16}
\end{equation}
sets an accuracy limit for refractivity measurement with the proposed method. 
With $\Delta {\mathcal E}/{\mathcal E} \approx 10^{-3}$ a systematic error for 
refractivity is $3.6\cdot 10^{-12}$ which corresponds to $\zeta= 7.3\cdot 10^{6}$
and an experimental reach to space crystal size of $1.2\cdot 10^{-28}m$.
Bending field inhomogeneity should contribute twice as less as  energy spread
since it enters to Eq.(\ref{horcor}) together with the energy and has $5 \cdot 10^{-4}$ 
relative uncertainty. 

\section {Cost estimate }

Since most of hardware should be adapted to existing beamline  
and magnets, there are no standard components available and 
therefor, we can only roughly estimate amount of necessary expenses.
Our estimates are shown in Table \ref{cost}.

\begin{table}[h]
\caption{Equipment expenses.}
\label{cost}
\begin{center}
\begin{tabular}{||l|l||}
\hline
Component & Cost (kEuro) \\ \hline\hline
Laser system & 250 \\ \hline
Beam pipe    &  75 \\ \hline
Septum magnet & 75 \\ \hline
Detectors    & 150 \\ \hline
\hline
%\multicolumn{2}{|c||}{ \includegraphics[scale=0.8]{PETRA_small} } \\
Total & 550 kEuro \\ \hline
\end{tabular}
\end{center}
\end{table}

Prices for optics and electronics are included in laser and detector costs 
respectively.

\newpage
\section {Conclusion }

A simple theoretical framework is established allowing to access extremely
small distances in laboratory, provided a vacuum refraction index growing with 
photon energy. Such vacuum is suggested by wide range of gravity theories 
which predict space-time modifications around Planck scale. 
Motivated by these predictions, we propose a laser Compton experiment at PETRA 
to test empty space for single or double refraction. Experiment would be able 
to prove or reject crystal-space hypothesis reaching distances as small as 
$10^{-28}m$ for refractivity and  $10^{-31}m$ for birefringence. Space isotropy
measurements within these magnitudes are also foreseen.

Space birefringence measurements would be performed with UV polarized
laser and would require 282 days of data taking (50\% efficiency assumed)
to reach $10^{-31}m$ sensitivity. Probing space isotropy within 
this running period is possible with a sensitivity $10^{-30}m$ by  
mapping $360^\circ$ celestial circle with $3.6^\circ$ steps.

For space refractivity tests one should explore an additional
green laser which will enable beam energy precise determination.
Very fast, sub-minute measurement times are sufficient to sample 
refractivity with an accuracy which corresponds to $10^{-28}m$ distance
sensitivity.   

%As a possible place of the experiment we have considered PETRA SWL
%and NWR locations. At SWL one can benefit using  existing infrastructure
%of Laser-Wire Compton experiment
 
Observation of either refractive or birefringent Planck space will have a large 
impact on gravity and related fields.
 
\section*{Acknowledgement}
We are gratefull to Reinhard Brinkmann for his support and advices related 
to the beam setup. One of the authors (V.G.) thanks Stefan Schmitt 
and Daniel Pitzl for useful discussions.

\end{document}